\magnification=\magstep2
\def\para{\par\noindent}
\def\sqr#1#2{{\vcenter{\vbox{\hrule height.#2pt
        \hbox{\vrule width.#2pt height#1pt \kern#1pt
          \vrule width.#2pt}
        \hrule height.#2pt}}}}

\newcount\notenumber

\def\note{\advance\notenumber by 1
\footnote{$^{\the\notenumber}$}}
\baselineskip 20pt
\centerline{{\bf Zero Temperature Dynamics of the}}
\centerline{{\bf  Weakly-Disordered Ising Model}}\para
\vskip 0.25cm
 \para S.~Jain,
 \para School of Mathematics and Computing,\para
 University of Derby,\para
 Kedleston Road,\para
 Derby DE22 1GB,\para
 U.K.\para
\vskip 0.25cm
\para E-mail: S.Jain@derby.ac.uk 
\vskip 3.0cm
\para Classification Numbers:
\para 05.20-y, 05.50+q, 05.70.Ln, 64.60.Cn, 75.10.Hk,
 75.40.Mg 
\vskip 1.5cm
\para
\para 
\vskip 1.0cm
\para
\para
\vfill\eject
\para {\bf ABSTRACT}
\para The Glauber dynamics of the pure and weakly-disordered random-bond
2d Ising model is
 studied at
zero-temperature. A single characteristic length scale, $L(t)$,
 is extracted from
the equal time correlation function. In the pure case, the persistence
 probability, $P(t)$, decreases algebraically with the coarsening
length scale.
\para In the disordered case, three distinct regimes are identified: a
short time regime where the behaviour is pure-like; an intermediate regime
where the persistence
 probability decays non-algebraically
with time; and
a long time regime where the domains freeze and there is a cessation of
growth.
 In the intermediate regime, we find that
$P(t)\sim L(t)^{-\theta^\prime}$, where $\theta^\prime = 0.420\pm 0.009$.
 The value of $\theta^\prime$
is consistent with that found for the pure 2d Ising model at zero-temperature.
\para Our results in the intermediate regime are
 consistent with a logarithmic decay
of the persistence probability with time, $P(t)\sim (\ln t)^{-\theta_d}$, 
where $\theta_d = 0.63\pm 0.01$.
\para 
\vfill\eject
\para The \lq persistence\rq\ problem is concerned with the determination of the
fraction of space which persists in the same phase up to some later time. So,
 for
spin systems we are interested in the fraction of spins that have not flipped
in some time $t$. This problem has been studied extensively over the last few
years [1-12] and, somewhat surprisingly, the persistence exponent 
($\theta$) has been found to
 be highly non-trivial
even for simple one-dimensional models, such as the $q-$state Potts model
 at zero temperature [6-7], of non-equilibrium coarsening dynamics. 
\para Although Stauffer [3] has performed Monte Carlo simulations in up to
5d, most of the work in higher dimensions has been largely limited to 2d.
 Numerical studies [1,3] estimate that
$\theta\sim 0.22 $ for the 2d Ising model with Glauber
dynamics at $T=0$. The analogous exponent for non-equilibrium critical
dynamics has also come under intensive investigation [9-12]. Very recently,
the persistence problem has been generalised to partial survivors [8].
 There has, however, been relatively
 little published in the literature
to date on the persistence problem in systems containing disorder. Here we
present the results of a numerical study of an Ising model containing quenched
impurities.    
\para In this work we study domain growth [13] in a weakly
disordered random-bond 2d Ising model and restrict ourselves to
 zero temperature.

\para The model we work with is given by
$${\it H} = -\sum_{<ij>} {J_{ij} S_i S_j}\eqno(1)$$
where the Ising spins ($S_i$) are assumed to be on
 every site of a square $N=500\times
500$ lattice with periodic boundary conditions and the summation runs
 over all nearest-neighbour pairs. The 
quenched ferromagnetic interactions are chosen from a binary distribution,
namely
$$P(J_{ij}) = (1-p)\delta(J_{ij}) + p \delta(J_{ij}-1)\eqno(2)$$
where $p$ is the concentration of bonds.
\para  The data presented here were
obtained on a suite of Silicon Graphics workstations.
\para  We work at
 zero temperature and consider a range of bond-concentrations in the vicinity
of the pure case, $0.975\le p\le 1.0$.
The initial configuration of the spins is chosen at random i.e. $S_i(t=0)=\pm 1
$ with equal probability for all $i$.
We then update the lattice using the following algorithm:
\item {1.} for a given spin $S_i$ we first 
calculate the local energy, $\Delta E_i$;
\item {2.} if $\Delta E_i < 0$, we leave $S_i$ as it is;
\item {3.} if $\Delta E_i = 0$, we flip $S_i$ at random (i.e.
 with a probability 1/2);
\item {4.} if $\Delta E_i > 0$, we flip $S_i$ with probability 1. 
\para We repeat steps 1 - 4 throughout
 the entire lattice during each Monte Carlo step.
\para  The number, $n(t)$,
of spins which have never flipped until time $t$ is then counted.
 In practice, we
record $n(t=t_r)$ where $t_r = 2^r, r=0,1,\dots,13$.
\para The persistence 
probability is defined by [1] 
$$P(t)=[<n(t)>]/N\eqno(3)$$
where $<\dots>$ denotes an average over different 
initial conditions and $[\dots]$ indicates an average over samples i.e.
 the bond-disorder; typically, {\it the number of different
 initial conditions}
$\times$ {\it the number of samples} = 100.    
\para During the simulations we also record the equal time pair correlation
 function, $C(r,t)$, which is
defined by [13]
$$C(r,t) = {1\over N}\sum_i[<S_i(t)S_{i+r}(t)>].\eqno(4)$$
According to the scaling hypothesis 
$$C(r,t)=f({r\over L(t)})\eqno(5)$$
where $f(x)$ is a scaling function and $L(t)$ is a single characteristic 
coarsening length scale.
 For the pure
model ($p = 1.0$) it\rq s now well established that $P(t)$
 decays algebraically [1]
$$P(t)\sim t^{-\theta}\eqno(6)$$
where $\theta\sim 0.22$. Furthermore, for the non-random
 model it\rq s also
well known that the domain length increases as $t^{1/2}$ [14]. 
\para Hence, from equation (6) we can write
$$P(t)\sim (t^{1/2})^{-\theta^\prime}\eqno(7)$$
where $\theta^\prime = 2\theta$.
\para We now turn to our numerical results. To begin with, we look at the pure
 ($p=1.0$) case to extract the value of $\theta^\prime$ mentioned above for
our model.
\para In Figure 1
we plot $\ln P(t)$ versus $\ln t^{1/2}$ for the pure case over
the time interval $2\le t\le 4096$.
The slope of the straight line gives $\theta^\prime = 0.418\pm 0.004$,
 which, of
course, implies that $\theta = 0.209\pm 0.002$, consistent with previous
results [1,3] (the error-bar quoted here is a statistical one).
\para As an independent check, we also extracted the coarsening length scale
by fitting the equal time correlation function, equation (4), to its expected
form, equation(5). Our results are completely consistent with
$$L(t)\sim t^{1/2}.\eqno(8)$$ 
\para When quenched impurities are introduced the domains grow more
slowly than in pure systems and for $T>0$ it\rq s expected that [15] they
increase as $ (T\ln t)^x $,
where the exponent $x=4$ in $d=2$ [16].
For $T=0$ we expect the quenched disorder to lead to an {\it eventual} cessation
of growth.

\para As we are working with weakly-disordered models, one would expect the
initial decay of $P(t)$ to be given by
equation (6). In Figure 2(a) we show a log-log plot of $P(t)$ against $t$
 for a range
of bond concentrations, $p: 0.975\le p\le 1.0$. Although the initial decrease
in $P(t)$ is indeed algebraic, there appears to be non-algebraic decay
before \lq freezing\rq\ sets in. This is shown explicitly in Figure 2(b)
where deviations from algebraic behaviour can be clearly seen. To investigate
this point further, we re-plot the data
in Figure 3 as $\ln P(t)$ against $\ln (\ln t)$. As a consequence, we see that
the persistence probability decays as
$$P(t)\sim (\ln t)^{-\theta_d}\eqno(9)$$
for a disordered system, where $\theta_d$ is now the persistence exponent,
before the long-time behaviour sets in.
 Furthermore, we notice that the behaviour
for the various disordered cases is qualitatively the same, irrespective of the
amount of disorder present.
\para For $p<1.0$ three distinct regimes can be identified: an initial short
 time regime ($t<t_1$) over which the behaviour is pure-like, an intermediate
regime ($t_1\le t\le t_2$) over which the persistence probability decreases
logarithmically and a final regime ($t>t_2$) where the system appears to
\lq freeze\rq\ and $P(t)$ effectively remains constant. It\rq s clear from 
Figure 3 that as disorder increases, $t_2$ decreases, i.e. the cessation
of domain growth is quickened by the strength of the disorder. The three
different regimes are clearly evident even in a very weakly disordered ($p=0.99$)
system. To ensure that we have a reasonably large intermediate regime to work
with, we now restrict our attention to the case where $p=0.99$. 

\para In Figure 4 we re-plot the data for $p=0.99$ over
 the range $16\le t\le 1024$.
The data for short times ($t < 16$) has been discarded as has the data over
 times ($t > 1024$) where the freezing of domains has occurred. The straight
line fit leads to a persistence exponent of $\theta_d = 0.63\pm 0.01$. This
result would appear to indicate a logarithmic growth of domains during the
intermediate regime at zero-temperature. This is somewhat surprising as 
the logarithmic behaviour discussed earlier is 
believed to hold true for {\it finite} temperatures.

 The behaviour of the growth before \lq freezing\rq\ sets in
 can be extracted independently
by fitting the equal time correlation function to its expected scaling form
given by equation (5). In Figure 5 we present the scaling plot of $C(r,t)$
for $p=0.99$. We plot $C(r,t)$ against $r/L(t)$,
 where $L(t)$ has been
chosen at each time to give the best data collapse. This clearly produces
 an excellent scaling plot. We stress that the plot shown in Figure 5 makes no
assumptions about the growth law of $L(t)$ with $t$. 
In Figure 6 we plot the data for the persistence probability for $p=0.99$
as a log-log plot of $P(t)$ against $L(t)$ where the latter has been extracted
from Figure 5. The linear fit in Figure 6 implies that $\theta^\prime=0.420\pm
0.009$, consistent with our earlier result for the {\it pure} case.
 Thus, expressing $P(t)$ in terms of the coarsening length scale leads to the
same behaviour as for the pure case.
\para To conclude, we have presented data for the zero-temperature
 dynamics of the
weakly disordered random-bond 2d Ising model. For the disordered system we
find evidence that $P(t)$ decreases logarithmically with 
time over an intermediate
regime. The
 (disordered) persistence exponent over this regime 
is estimated to be $\theta_d=0.63\pm 0.01$.
However, for both the pure and the disordered models the persistence
probability is found to decay
algebraically with the coarsening length scale with the {\it same} exponent.        
At present we are studying generalised persistence [8-12] for disordered models.
\para {\bf Acknowledgement}
\para I would like to thank Alan J Bray for useful correspondence
 during the initial
stages of this work and for a critical reading of the draft manuscript.
 Matthew Birkin is also thanked
 for both technical assistance and maintaining the Silicon
Graphics workstations.
\vfill\eject 
\para FIGURE CAPTIONS
\vskip 1cm
\para Figure 1
\para A log-log plot of $P(t)$ against $t^{1/2}$, for
the pure 2d Ising model. The linear fit shown gives a value of $\theta^\prime
= 0.418\pm 0.004$.
\vskip 1cm

\para Figure 2(a)
\para A plot of $\ln P(t)$ against $\ln t$ for a range of
 bond-concentrations, $p$; the data for the pure case, $p=1.0$, is plotted
for comparison and the straight line has slope $-0.209$.
\vskip 1cm

\para Figure 2(b)
\para A re-plot of some of the data shown in Figure 2(a) on an expanded
scale to highlight the deviations from algebraic decay; the linear fit is
for the pure case with $\theta = 0.209$.
\vskip 1cm

\para Figure 3
\para A plot of $\ln P(t)$ against $\ln (\ln t)$ for a range
 of bond-concentrations, $p$; the data for the pure case, $p=1.0$, is plotted
for comparison.
\vskip 1cm

\para Figure 4
\para A re-plot of the data for $p=0.99$ from Figure 3. The straight line
fit confirm a logarithmic decay of the persistence probability over time.
The gradient of the line shown leads to $\theta_d = 0.63\pm 0.01$.
\vskip 1cm

\para Figure 5
\para A scaling plot of $C(r,t)$ versus $r/L(t)$, where $L(t)$ has been chosen
at each time to give the best data collapse.
\vskip 1cm

\para Figure 6
\para A plot of $\ln P(t)$ versus $\ln L(t)$, where $L(t)$ has been extracted
from Figure 5.
The linear fit confirms that the persistence probability decays algebraically
with the coarsening length scale, $L(t)$. The slope of the straight line yields
$\theta^\prime = 0.420\pm 0.009$.
\vfill\eject
\para REFERENCES
\item {[1]} Derrida B, Bray AJ and Godreche C 1994 J.Phys.A: Math. Gen.
 {\bf 27} L357
\item {[2]} Bray AJ, Derrida B and Godreche C 1994 Europhys. Lett. {\bf 27} 177
\item {[3]} Stauffer D 1994 J.Phys.A: Math. Gen. {\bf 27} 5029
\item {[4]} Dornic I and Godreche C 1998 J.Phys.A: Math. Gen. {\bf 31} 5413
\item {[5]} Baldassarri A, Bouchaud JP, Dornic I and Godreche C 1998
 Cond-mat/9805212
\item {[6]} Derrida B 1995 J.Phys.A: Math. Gen. {\bf 28} 1481
\item {[7]} Derrida B, Hakim V and Pasquier V 1995 Phys. Rev. Lett. {\bf 75}
751; 1996 J. Stat. Phys. {\bf 85} 763
\item {[8]} Majumdar SN and Bray AJ 1998 Phys. Rev. Lett. {\bf 81} 2626
\item {[9]} Majumdar SN, Bray AJ, Cornell SJ and Sire C 1996 Phys. Rev. Lett.
{\bf 77} 3704
\item {[10]} Oerding K, Cornell SJ and Bray AJ 1997 Phys. Rev. E{\bf 56} R25
\item {[11]} Zheng B 1998 Int. J. Mod. Phys. B{\bf 12} 1419
\item {[12]} Drouffe J-M and Godreche C 1998 Cond-mat/9808153
\item {[13]} Bray AJ 1994 Adv. Phys. {\bf 43} 357
\item {[14]} Gunton JD, San Miguel M and Sahni PS 1983 {\it Phase Transitions
and Critical Phenomena} vol 8 ed C Domb and JL Lebowitz (New York: Academic
Press)
\item {[15]} Huse DA and Henley CL 1985 Phys. Rev. Lett. {\bf 54} 2708 
\item {[16]} Huse DA, Henley CL and Fisher DS 1985 Phys. Rev. Lett. {\bf 5}
1924
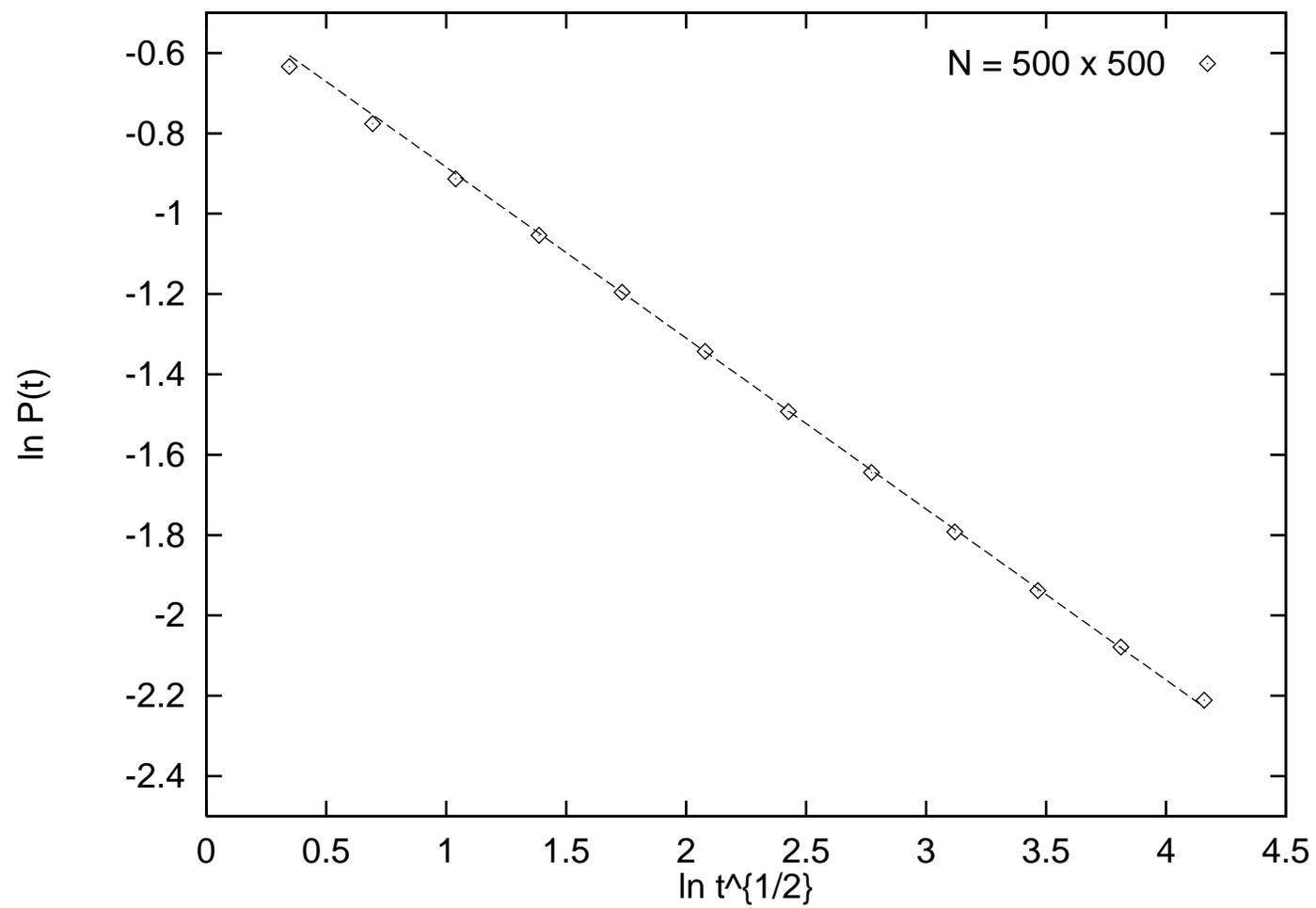
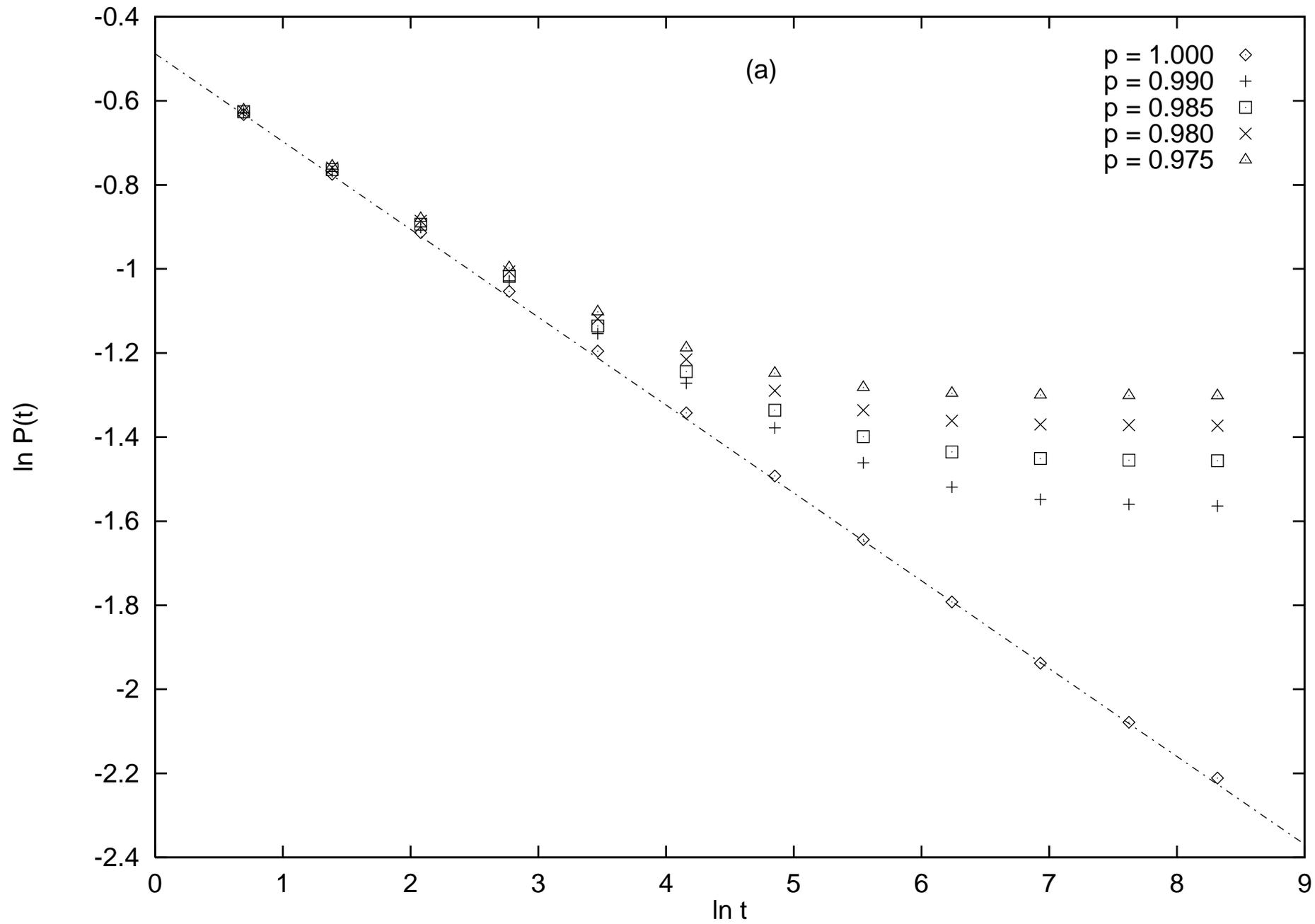
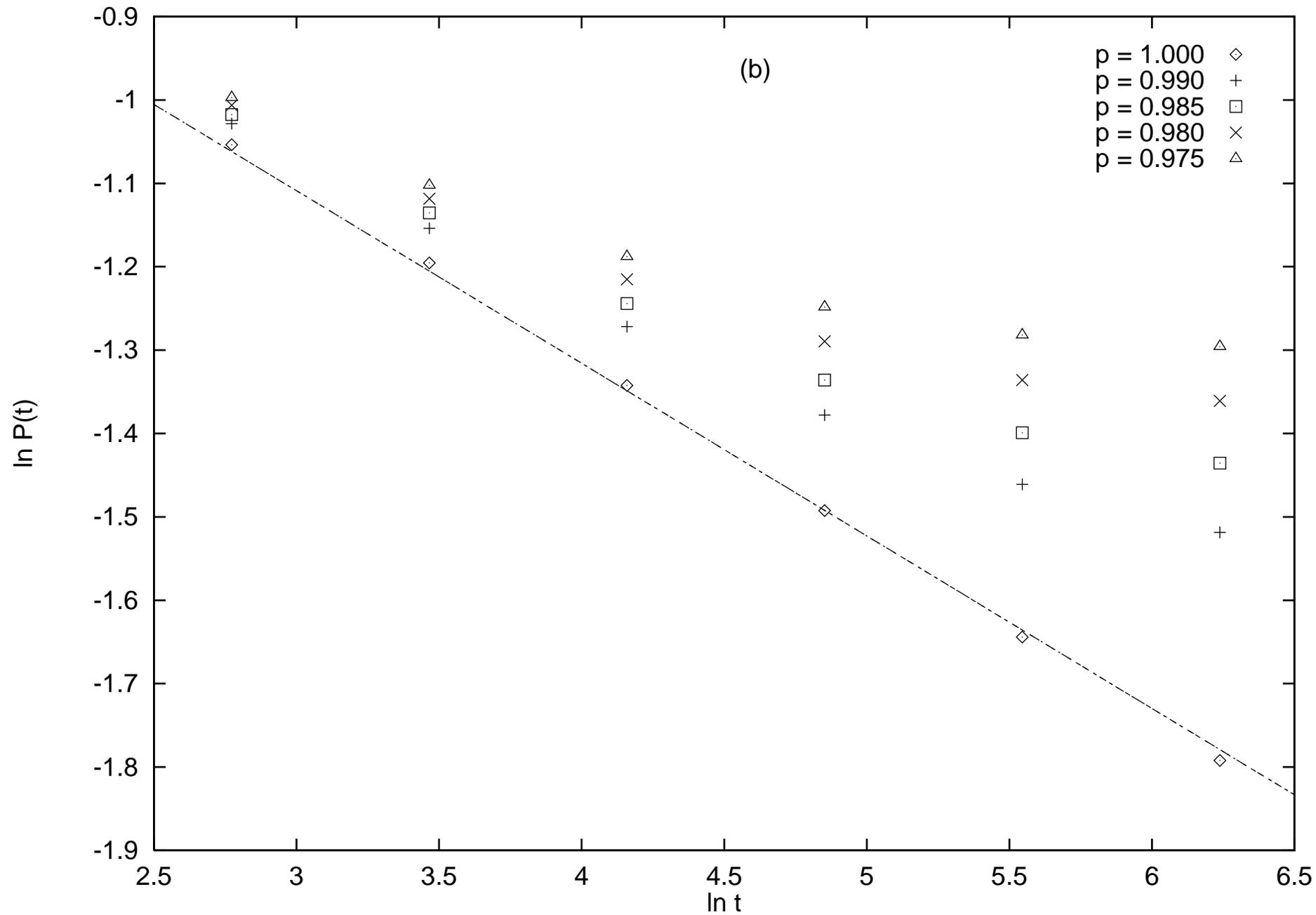
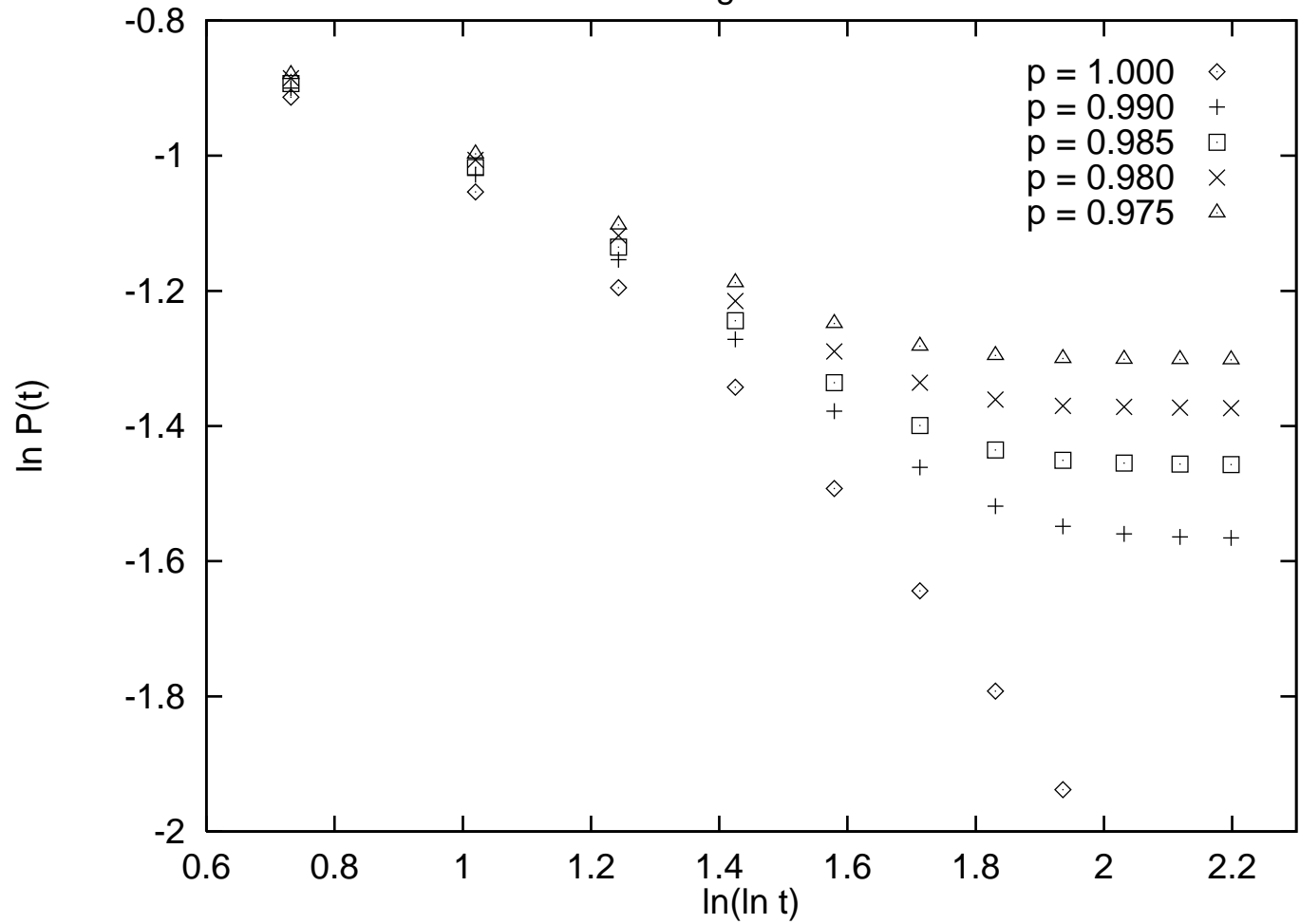
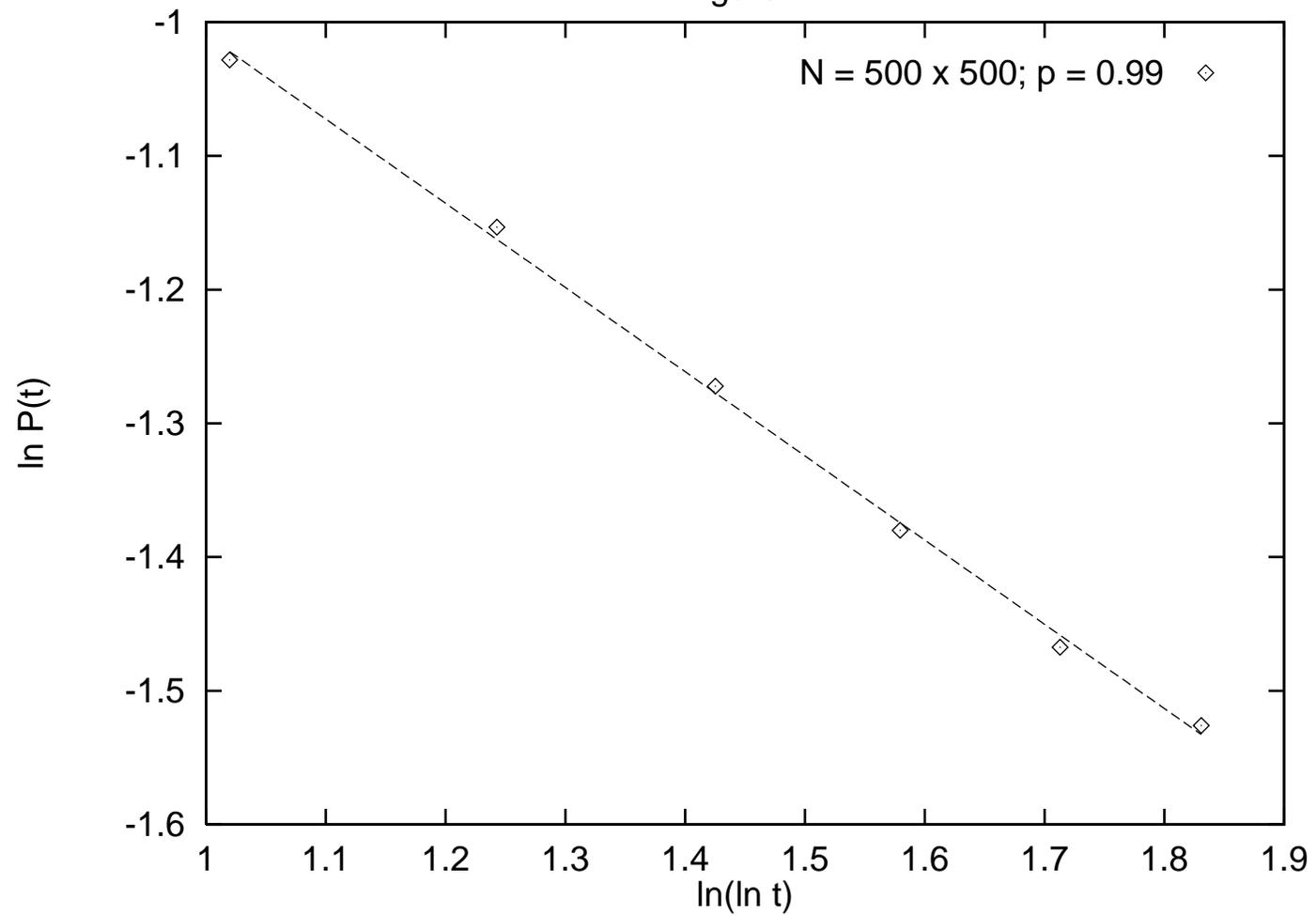
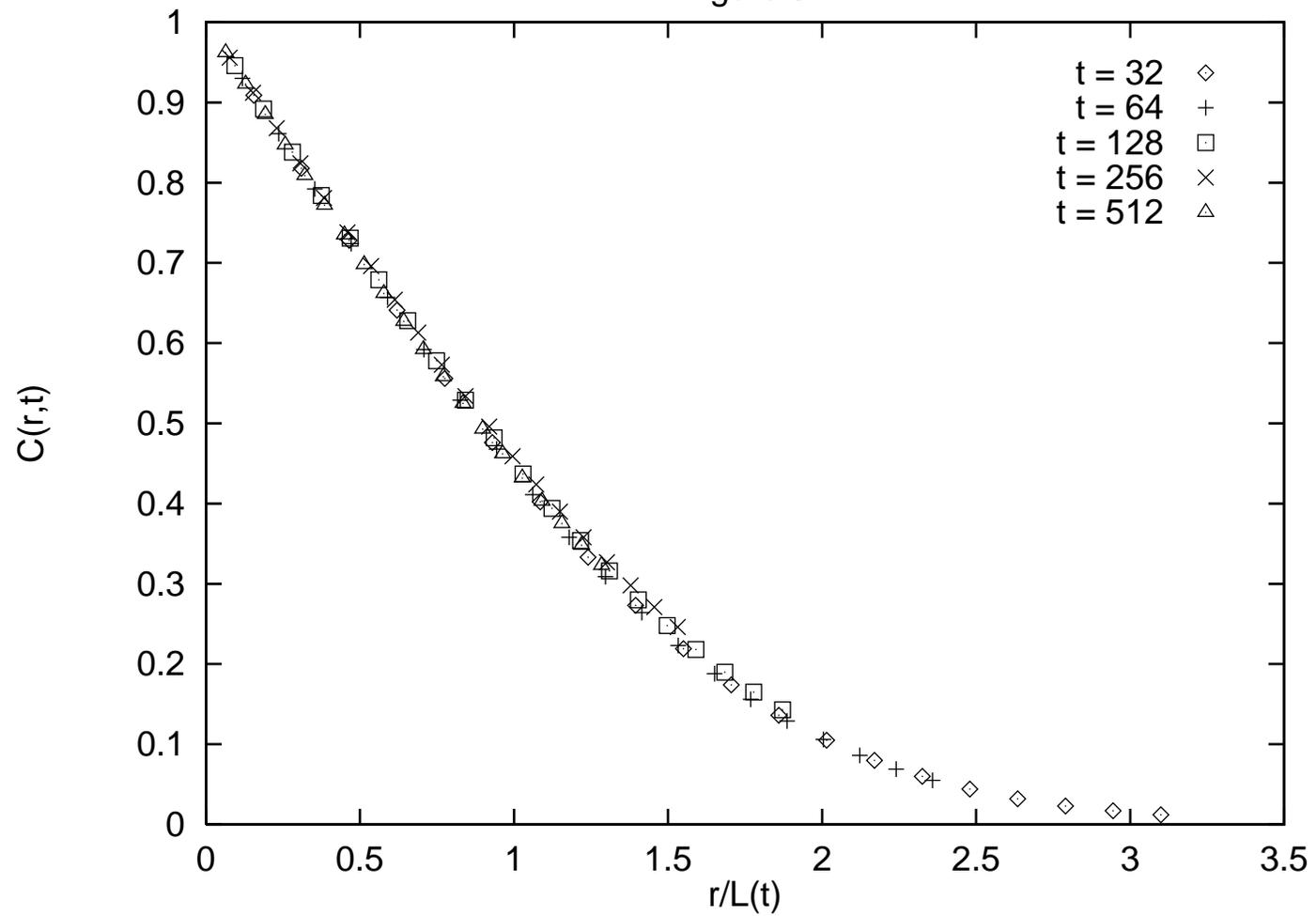
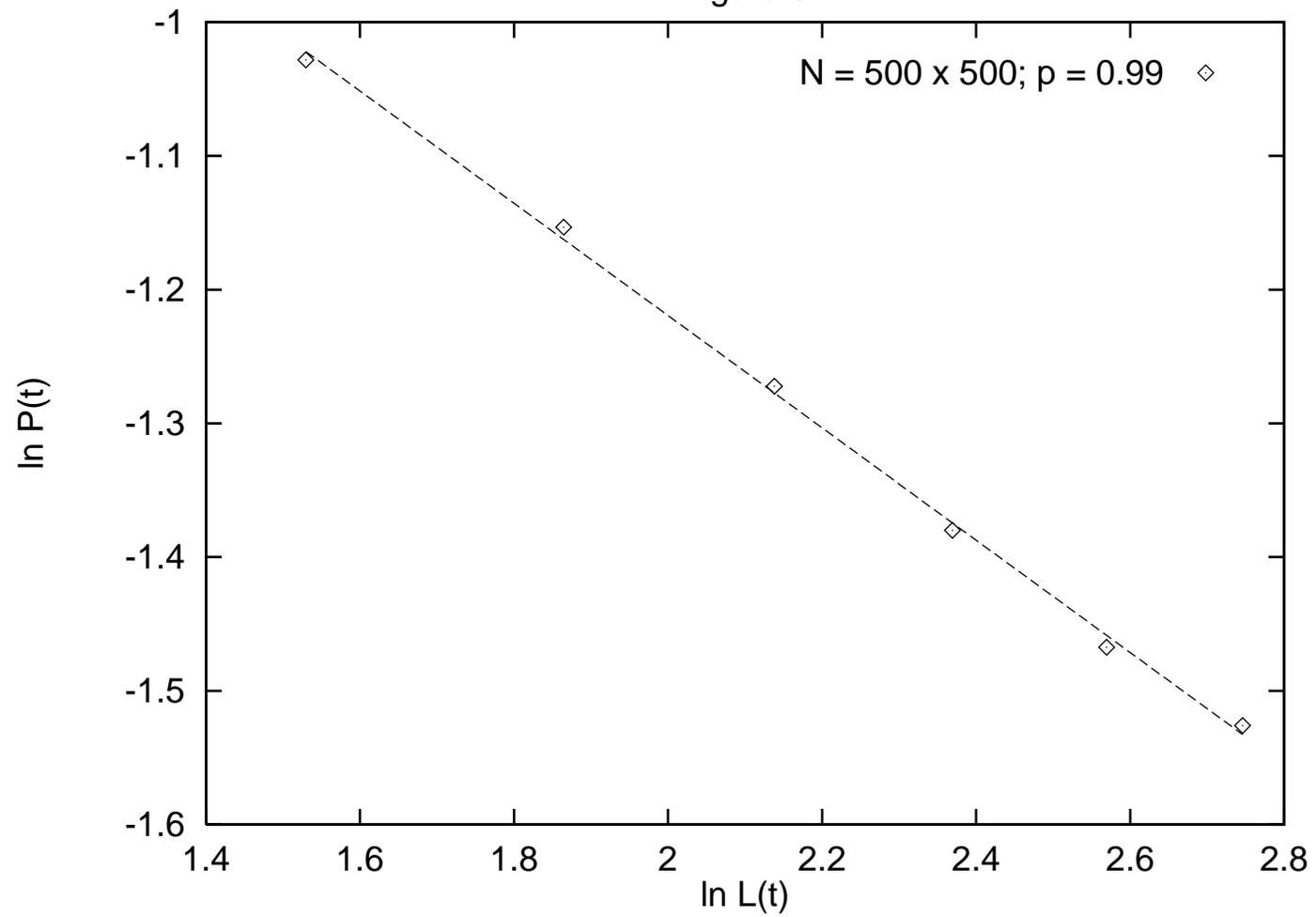
\end